\documentclass[12pt]{article}
\usepackage{amsmath}

\def\qed{\relax\ifmmode\hskip2em \fbox{ }\else\unskip\nobreak\hskip1em 
$\fbox{}$\fi}

\newsavebox{\theorembox}
\newsavebox{\lemmabox}
\newsavebox{\corollarybox}
\newsavebox{\propositionbox}
\newsavebox{\examplebox}
\newsavebox{\propertybox}
\savebox{\theorembox}{\bf Theorem}
\savebox{\lemmabox}{\bf Lemma}
\savebox{\corollarybox}{\bf Corollary}
\savebox{\propositionbox}{\bf Proposition}
\savebox{\examplebox}{\bf Example}
\savebox{\propertybox}{\bf Property}
\newtheorem{theorem}{\usebox{\theorembox}}

\newtheorem{example}{\usebox{\examplebox}}
\newtheorem{definition}{{\sc Definition}\rm }[section]
\newtheorem{definitions}[definition]{{\sc Definitions\rm }}

\newlength{\proofskip}
\begin{document}

\begin{center}

{\LARGE 
A Fast Scalable Heuristic for Bin Packing
}


\footnotesize

\mbox{\large Srikrishnan Divakaran}\\
School of Engineering and Applied Sciences, Ahmedabad University, 
Ahmedabad, Gujarat, India 380 009,
\mbox{
Srikrishnan.divakaran@ahduni.edu.in }\\[6pt]
\normalsize
\end{center}

\baselineskip 20pt plus .3pt minus .1pt


\noindent 
\begin{abstract}
In    this paper we   present a fast scalable heuristic   for  bin   packing that partitions the given problem into identical  sub-problems of       constant size and solves these constant size 
sub-problems by considering only a constant    number  of bin configurations with bounded unused space. We  present some   empirical evidence to   support the scalability  of our heuristic  and its tighter empirical analysis of hard instances       due to   improved   lower    bound on the necessary wastage in an optimal solution.
\end{abstract}
\bigskip
\noindent {\it Key words:}  
Keywords: Bin Packing; Cutting Stock Problems; Heuristics; Approximation Algorithms; Approximation Schemes; PTAS; Exact Algorithms; Design and Analysis of Algorithms.
\noindent\hrulefill
\section{Introduction}
The {\em Bin Packing} problem is a classical combinatorial  optimization problem that has been widely studied since the 1970's and can be stated as  follows: 
\begin{quote}
Given a collection $\cal{B}$ of unit capacity   bins and a    sequence $L= (a_1,a_2, ...,a_n)$   of  $n$  items  with their respective sizes $(s_1,s_2,...,s_n)$ such that $\forall{i}$ $s_i\in 
[0, 1]$,      determine a packing of the items in $L$ that uses a minimum number  of bins from $\cal{B}$. 
\end{quote}
This problem has a wide variety of applications\cite{JDUGG74} including cutting stock applications, packing            problems in supply chain management, and resource  allocation problems in   distributed systems. Algorithms     for bin packing attempt to pack the items   in $L$ using minimum number of bins in $\cal{B}$ and can be broadly    classified as {\em offline} and {\em online}. Offline   algorithms     are algorithms that pack items with complete knowledge of the list $L$ of items  prior to packing, whereas online algorithms need to pack items as they arrive without any knowledge of future. The bin packing  problem even  for the offline  version  is  known  to  be   NP-Hard\cite{GJ79} and hence most research efforts have focused on the design of fast  online and offline approximation algorithms with good performance. The   performance of an approximation algorithm is defined  in terms of its worst case behavior  as follows: Let  $A$ be an algorithm for  bin packing and    let $A(L)$ denote the number of bins required     by $A$ to pack items in $L$, and OPT denote          the optimal algorithm for packing items in $L$. Let   $\cal{L}$ denote the set of all possible list sequences whose items are of sizes in $[0, 1]$. For         every $k > 1$, $R_{A}(k) = sup_{L \in \cal{L}}^{} \{A(L)/k : OPT(L) = k \}$. Then the asymptotic worst   case  ratio is given by $R_{A}^{\infty} = lim_{k\rightarrow \infty}R_{A}(k)$.  This ratio    is the asymptotic approximation ratio  and measures   the quality of the algorithms packing in comparison  to the optimal packing in the worst case scenario. The second way of measuring the performance  of an approximation algorithm     is  $sup_{L \in \cal{L}}^{} \{ A(L) / OPT(L)\}$ and this ratio is the absolute approximation ratio     of the algorithm. In         the case of online algorithms this ratio is often referred to as competitive ratio. A polynomial time approximation scheme (PTAS) for bin packing is a class of algorithms  that given any instance $I$ and an $\epsilon\in (0,1)$ produces an approximation algorithm whose solution    quality is within $(1+\epsilon)$ times the optimal  solution quality and         its computational time is polynomial in its size and $\frac{1}{\epsilon}$. A PTAS    essentially  allows  the  user to choose by specifing the parameter $\epsilon$ the  algorithm     from this class that guarantees an $(1+\epsilon)$ optimal solution.
However, stricter performance guarantee  is achieved by these algorithms at the cost of increased computation. The computation cost in   practice is very high even for $\epsilon \approx 0.1$.  
The  high computational cost of    PTAS, coupled with   the inability to  provide strict theoretical guarantees for many algorithms   that perform well in practice, and        traditional worst case performance measures categorizing an algorithm's performance as poor based        on very few degenerate instances           has motivated the need for {\em efficient  heuristics - fast algorithms  that perform well on most instances without requiring any theoretical guarantees on its worst case behavior}. In      addition, the phenomenal growth in volume of data has driven the need for      heuristics that can scale computationally and are amenable to  tighter empirical analysis.  \newline \newline
In this paper we employ simple combinatorial      ideas in    the  design and analysis of a fast scalable heuristic  that partitions the given   problem into identical sub-problems of constant size and solves these constant size sub-problems by considering only a constant   number of bin configurations with bounded unused space. We present our empirical study that provides evidence for the scalability  of our heuristic and its tighter  empirical analysis   for  hard instances due to improved   lower    bound on the necessary wastage in an optimal solution.
\section{Related Results}
In this section, we summarize the       main results in bin packing from the perspective of         approximation algorithms and  polynomial time approximation schemes. For a detailed survey of these and other  related results, we refer the reader to Johnson's Phd Thesis \cite{J73}, Coffman 
et al.\cite{CCGMV13} and Hochbaum\cite{H97}.   
\newline 
{\bf Online Algorithms}:
$\text{NEXT-FIT(NF)}$, $\text{FIRST-FIT(FF)}$ and   $\text{BEST-FIT(BF)}$ are the most widely studied natural and classical online   algorithms for bin packing. Johnson et al.\cite{J73,J74, JDUGG74} showed that  both $FF$ and $BF$ have an asymptotic competitive ratio of $1.7$. Subsequently, Yao presented an online algorithm  $\text{REVISED-FF(RFF)}$\cite{ Y80}  based on $FF$ that achieved  an  asymptotic competitive  ratio of $5/3$.   This was further improved by     Lee and Lee\cite{LL85}, Seiden\cite{S02}, and more recently Balogh et al.\cite{ BBDSV15} settled    this online problem by presenting   an optimal online    bin packing with absolute worst case competitive  ratio of $5/3$. \newline 
{\bf Offline Algorithms}:
The most natural offline algorithms for bin packing essentially   reorder the items    and then employ other classical online algorithms like $NF$, $FF$, $BF$ or         other online algorithms to pack the items. This has resulted in three simple but effective       offline algorithms; they are denoted by $NFD$, $FFD$,             and $BFD$, with the “D” standing for “Decreasing". The sorting needs $O(nlog n)$ time and so the total running time of each of these algorithms is $O(nlogn)$. Baker and   Coffman \cite {BC81} established     the asymptotic approximation ratio for $NFD$ to be $\approx 1.69103$,     Johnson et al.\cite{JDUGG74}, Baker \cite{B85} and Yue\cite{Y91} established $FFD$ and $BFD$'s aymptotic approximation ratio to be $11/9$. Subsequently, $\text{ Refined-First-Fit Decreasing (RFFD)}$ by Yao $\cite{Y80}$, Modified First Fit (MFFD) by       Garey and Johnson \cite{GJ85}, $Best-Two-Fit (B2F)$      and $Combined Algorithm (CFB)$  by Friesen and Langsten\cite{FL91}       resulted in achieving an asymptotic competitive ratio of $\approx 1.18$      but at a very high computational cost. \newline 
{\bf Approximation Schemes}:                  Fernandez    de la Vega and Lueker\cite{FL81} designed a PTAS that for any real number    $\epsilon > 0$, constructed a $(1+ \epsilon)$ optimal  solution in $C_{\epsilon} +   Cnlog(1/\epsilon)$       time where $C_{\epsilon}$ and $C$ are      large constants that depend on $\frac{1}{\epsilon}$.  Subsequently, Johnson\cite{ J82}, Karmarkar and   Karp\cite{ KK82} and     Hochbaum   and Shmoys\cite{ HS86, HS87}  presented improved approximation schemes. These  approximation  schemes helped in  obtaining near optimal solutions with its computation time polynomial in its   size and  $\frac{1}{\epsilon}$. For $\epsilon < 0.1$, the   computational time even            for moderate sized instances made these PTAS  practically not usable. This   high computational cost of 
PTAS coupled with the inability to  provide strict theoretical guarantees for many algorithms that perform well         in practice has lead to the study of heuristics. In this 
paper our focus is on heuristics based on simple combinatorial ideas (simple and 
effective heuristics using  combinatorial ideas are mostly similar to the online and offline algorithms already described earlier in this section) and hence
we do not go into the  heuristics based on approaches like branch and bound, local search, simulated annealing, tabu search, genetic algorithms  and constraint optimization. However, the interested reader can look at the survey paper of Delorme et al\cite{DIM16} for 
these results. 
\subsection{Our Results}
In this paper, we first present Algorithm $B(\epsilon)$, an algorithm that  given a real valued parameter $\epsilon \in (0, \frac{1}{2})$, partitions the  original problem into many identical  sub-problems of size $c\in [\lceil\frac{1}{\epsilon}\rceil,\lceil\frac{2}{\epsilon}\rceil]$ and then uses  exact algorithms or existing PTAS to   solve these $c$-length  bin packing problems. Then we present Heuristic $C$, a heuristic that just like Algorithm $B$ partitions the original problem        into may identical $c$-length problems but solves the $c$-length sub-problems by considering only a constant number of bin  configurations with wastage very close to $\epsilon$ (i.e. an     extremely     small fraction of the bin configurations considered by PTAS or exact algorithms). This  results in significant reduction in computation time without any   noticable impact on its performance guarantee. Finally, we conducted an empirical study of Heuristic  $C$ involving         several hundred large instances of both    randomly generated as well as hard  instances to study its       computational scalability under the constraint that it provides an approximation                guarantee of $1.1$. For most of    the instances Heuristic $C$ was computationally  scalable (i.e. the  problem instance were split into identical sub-problems of size $c<10$ which were  then solved by considering less than $10$ distinct bin configurations). For  some instances Heuristic $C$  needs to consider $\approx 25$ distinct bin   configurations 
in order       to satisfy the performance guarantee constraint. For some  instances traditional analysis did not establish the desired   performance guarantee, but we were able to  obtain the desired performance guarantee by     obtaining a better lower bound on the necessary wastage in
 an optimal solution. The rest of this paper is organized as follows: 
In Section $3$    we    present       Algorithm $B$, in Section $4$ we 
present Heuristic $C$, and in Section $5$ we present our empirical 
study of Heuristic $C$.
\section{Bin Packing Based on Near Identical Partitioning}
In this section, we   present    an algorithm that given a real valued 
parameter        $\epsilon \in (0, \frac{1}{2})$, partitions the input sequence $L$ into  identical    sub-sequences     (except for the last sub-sequence) of   length   $c \in [\lceil \frac {1}{\epsilon} \rceil, \lceil\frac{2}{\epsilon}\rceil]$ (i.e. sum  of sizes of items in these subsequences is $c$) and then makes use of either  exact algorithms or known PTAS to  pack  the items in these $c$-length  subsequences  onto  unit    capacity       bins. We now introduce some necessary terms and definitions, before presenting our algorithm and its analysis. 
\begin{definitions} 
	The sequence $L=(a_1,a_2, ..., a_n)$   with $k$ distinct   item sizes    \{$s_1, s_2, ..., s_k$\}       can       be       viewed as    a  $k$    dimensional vector $d(L) = (n_1*s_1,n_2 *s_2,...,n_k*s_k)$, where for  $i \in [1..k]$, $n_i$ is the number of items of      type  $i$ (size $s_i$); we             refer to $d(L)$  as  the distribution vector corresponding to $L$. 
	For a   given real number $c > 1$, 
	let  $d_{c}(L)$  denote a  $c$-length segment  of $d(L)$ (i.e. a vector that is parallel to $d(L)$ and contains its initial segment such that its component  sum equals   $c$), and 
	$\text{min-packing}(d_{c}(L))$ to be the smallest sized bin packing of items corresponding to $d_{c}(L)$.
\end{definitions} 
{\bf Remark} : 
If the  number of   distinct sizes in $L$ is not  bounded  by a constant $k$, then we can still apply the above idea by partitioning the interval  $[0,1]$ into     $k$ distinct     sizes   $0, 1/k, 2/k, ..., 1$ and    round  the   item  sizes  in  $L$   to  the nearest     multiple  of $1/k$ that is   greater         than or  equal to the item size.  
\newline \newline  
{\bf Key Idea}: For an integer   $c^{*}\in [\lceil \frac{1} {\epsilon} \rceil,\lceil\frac{2}{\epsilon}\rceil]$, we partition the  distribution vector $d(L)$ into many copies of $d_{c^{*}}(L)$, the $c^{*}$ length initial segment of $d(L)$ (except for the last segment), where          $c^{*}$ is determined as follows:
For each   $c\in [\lceil\frac{1}{\epsilon}\rceil,\lceil\frac{2}{\epsilon} \rceil]$, we   determine the  packing ratio  $\frac{\text{min-packing}(d_{c}(L))}{c}$. Then, we choose $c^{*}$ to be $c \in [\lceil\frac{1}{\epsilon} \rceil,\lceil
\frac{2}{\epsilon}\rceil]$ for   which the packing ratio is minimum.
\begin{example}
Let    us consider a sequence $L$ of $3000$ items consisting of $600$ items of size  $0.52$, $600$ items of size $0.29$, $600$     items of size $0.27$ and $1200$ items of  size $0.21$. Let $\epsilon = 0.1$ is the approximation ratio desired.  For  this instance the distribution vector $d(L)$ is a      $4$-dimensional vector $(0.21*1200, 0.27*600, 0.29*600,0.52*600)=(252,162,174,312)$ of  length $900$. Our algorithm   attempts to  partition $d(L)$ into a $c$-segment vector for some  $c$  between $(1, \lceil \frac{2} {\epsilon} \rceil)$. We can observe that we can partition $d(L)$ into $60$ copies of the segment vector $(4.2, 2.7, 2.9,5.2)=(0.21*20, 0.27*10,0.29*10, 0.52*10)$ of    length $15$. 
The  $\text{min-packing}$ for this  segment vector of length  $15$  can be  determined using any of the exact algorithms or     existing PTAS 
for regular bin packing with $\epsilon=0.1$.
\end{example}
\begin{example}
Let us consider a sequence $L$ of $3000$ items consisting of $1000$ items of size  $0.60$, $1000$ items of size $0.65$,      and $1000$ items of     size $0.75$.
Let us consider the problem instance in Example $1$  with $\epsilon = 0.1$ is     the   approximation ratio desired.  For this instance the distribution   vector $d(L)$ is a $3$-dimensional vector $(0.60*1000, 0.65 *1000, 0.75*1000)=(600, 650, 750)$        of  length $2000$. Our  algorithm    attempts to  partition $d(L)$ into         a $c$-segment vector for some  $c$ between $(1,  \lceil \frac{2}{\epsilon}\rceil)$. We can observe that  we can partition $d(L)$ into $100$ copies of the segment vector $(6.0,6.50, 7.50)=(0.60*10,0.65*10,0.75*10)$ of length $20$. The  $\text{min-packing}$ for this segment vector of length $20$ can be determined using exact algorithms or any of the existing PTAS for regular bin packing with $\epsilon=0.1$.
\end{example} 
{\bf ALGORITHM B($L$, $\epsilon$)}
\begin{tabbing}
	Input(s): \=  (1) \= $L$ \  = \=  $(a_1,a_2,...,a_n)$ \  be the sequence of $n$ items with  their     respective sizes  \\ 
	\>      \>         \> $(s_1,s_2, ... ,s_n)$ in the interval $[0, 1]$; \\
	\> (2) \> $\epsilon \in (0, \frac{1}{2})$ be a user specified parameter; \\
	Output(s): The assignment of the items in $L$ to the bins in $\cal{B}$; \\
	Begin \= \\ 
	\> (1) \= Let \= $d(L)=(s_1*n_1,s_2*n_2,...,s_k*n_k)$  be the distribution vector corresponding to $L$; \\
	\> (2) \> For\=\ ($c=\lceil \frac{1}{\epsilon} \rceil$; $c\le\lceil \frac{2}{\epsilon}\rceil$; $c= c+1$) \\
	\> (2a)\>     \> Let \= $d_{c}(L)=(s_1*n^{c}_1,s_2*n^{c}_2,...,s_k*n^{c}_k)$ be the $c$-length 
	\ initial\  segment\  of\  $d(L)$  \\
	\>     \>     \> \>  with a packing ratio  $\text{packing-ratio}(c)=\frac{\text{min-packing} (d_{c}(L))}{c}$; \\
	\> (3) \> Let $c^{*}$\= be an integer in $(\lceil \frac{1}{ \epsilon }\rceil, \lceil \frac{2}{ \epsilon }\rceil)$ :  $\text{packing-ratio}(c^{*}) = \min_{c\in (\lceil \frac{1}{\epsilon}\rceil,\lceil \frac{2}{\epsilon}\rceil)}^{} \text{packing-ratio}(c)$;\\
	\> (4) \> Let $T= d_{c^{*}}(L)$ and $l = \frac{|d(L)|}{c^{*}}$; \\
	\> (5) \> return  $\bigcup \limits_{i=1}^{l} \text{min-packing}(T) \cup \text{min-packing}(d(L)- l*T)$; \\
	End 
\end{tabbing}
\begin{definitions}
Let Algorithm $B(L, \epsilon)$ partition $d(L)$ into $l$ copies 
of $T = d_{c}(L)= (n^{c}_1*s_1, n^{c}_2*s_2,...,n^{c}_k*s_k)$ (discarding the last segment), 
where $c$ is an integer in $[1, \lceil\frac{2} {\epsilon}\rceil]$ and
$T$ is a $c$-length initial segment of $d(L)$.
Let $T^{t}=        (\lfloor n^{c'}_1 * s_1 \rfloor , \lfloor n^{c'}_2   *s_2 \rfloor, ...,  \lfloor n^{c'}_k * s_k \rfloor)$  be the segment vector  obtained by truncating for $i \in [1..k]$, the $i$th components of $T$ to the nearest integer multiple of $s_i$.
Let $\text{Packing}(d(L))$ be the $\delta$-packing determined by Algorithm $B$ for $d(L)$. 
\end{definitions}
\begin{theorem}
	If (i) $T = T^{t}$ (i.e. for $i \in [1..k]$ the $i$th component is an integer multiple of $s_i$) ; 
	OR (ii) $\sum_{i=1}^{k} s_i = o(c)$ OR $k =o(c)$, then Algorithm $B$ constructs an asymptotically optimal packing for $d(L)$.
\end{theorem}
\section{A Fast Heuristic Based on Near Identical Partitioning}
The Algorithm $B$                           constructed the bin packing for                   sequence $L$ by   essentially  partitioning   the distribution      vector $d(L)$         into $l$ identical copies  of a $c$-length         segment $d_{c}(L)$     (except for the last segment) and      then constructing    $\text{min-packing}$ of    $d_{c}(L)$ using either exact methods or known PTAS for bin packing. However these exact algorithms or PTAS construct  $\text{min-packing}(d_{c}(L))$     by considering all  possible bin configurations  of unit capacity bins and hence are  computationally expensive. We    address this  computational issue by designing a Heuristic $C$ that constructs    $\text{min-packing} (d_{c}(L))$ by restricting the choice of vectors (bin configurations) to a small subset of $(1-\delta)$-vectors (i.e. vectors that correspond to bin configurations with unused space of at   most $\delta$).    This      restriction         results in significant   improvement   in the  computational efficiency of the Algorithm $B$              without significant downside on its solution quality. 
Also, for  many hard instances of bin packing   an optimal bin packing  is       not compact because of large unavoidable wastage in their bin packing. This             wastage in   an optimal solution    is often underestimated resulting in weak analysis on the performance of PTAS / approximation algorithms. We   address this analysis  problem by using $(1-\delta)$ vectors to get a   better  lower bound on the wastage  in an optimal bin packing.  
We now introduce some definitions necessary  for describing  Heuristic $C$. Heuristic $C$ will   make use of a sub-routine $\text{min-packing}_
{ \delta}$ that will be defined subsequently.
\begin{definitions}
For a real number $\delta \in (0,\frac{1}{2})$, the configuration of a unit capacity bin containing items whose sizes are from  $\{ s_1, s_2, ..., s_k \}$ and has a  wastage of at most $\delta$ can be   specified by a $k$-dimensional vector whose $i^{th}$ component, for $i\in [1..k] $, is the sum of sizes of  items of type $i$ (size $s_i$) in that bin; and its length is    in the interval $[1-\delta, 1]$,  where  the length of      a           vector  is defined  to be the sum of  its components.  We refer to  such  a  vector  as a $(1 - \delta)$-vector (bin configuration)  consistent              with $L$; and we denote by $e_{\delta}(L)$ the set of all  ($1-\delta$)-vectors (bin configurations) consistent with $L$. 
\end{definitions} 
\begin{definitions}
	For a given sequence $L$ and a real number $\delta\in (0,1/2]$, if $e_{\delta}(L)$ is non-empty then we define
	\begin{itemize} 
		\item [-]  a  $\delta$-packing for $d(L)$ to be  a minimal collection of  $(1-\delta)$-vectors   from $e_{\delta}(L)$   such that for $i \in [1..k]$, the sum of the $i$th component of these collection  of vectors  is greater than or equal to the $i$th component of   $d(L)$; 
		\item [-] $\text{min-packing}_{\delta}(d(L))$ to be a $\delta$-packing for $d(L)$ of the smallest size; If 
		for a given $\delta$, if it is not possible to pack 
		$d(L)$ using vectors from $e_{\delta}(L)$ then 
		$|\text{min-packing}_{\delta}(d(L))| = \infty$.
\end{itemize} 
\end{definitions}
Note: For certain sequences $L$, the item sizes in $L$  may be such that for certain values of $\delta \in (0, \frac{1}{2})$ there are no $1-\delta$ vectors consistent with $L$ (i.e. $e_{\delta}(L)$ is empty). 
\begin{example}
	Let us consider the problem instance in Example $2$  with $\epsilon = 0.1$ is     the   approximation ratio desired.  For this instance the distribution vector $d(L)$ is a $3$-dimensional vector $(0.60*1000,0.65*1000, 0.75*1000 )=(600, 650,750)$ of length $2000$. For this instance if  $\delta < 0.25$ then there are no $1-\delta$ vectors consistent with $L$ and for $\delta < 0.4$ there are no 
	$\delta$-packings of $L$.
\end{example}
{\bf Key Idea}: For  an   integer $c^{*}\in [\lceil \frac{1}{\epsilon} \rceil,    \lceil     \frac{2}{\epsilon} \rceil]$, we   partition  the distribution     vector $d(L)$ into many copies of $d_{c^{*}}(L)$, the $c^{*}$ length segment of $d(L)$ (except for  the last segment), where $c^{*}$ is determined as follows:
For each $c\in [\lceil\frac{1}{\epsilon}\rceil,\lceil\frac{2}{\epsilon
}\rceil]$, we determine $\delta_{c}$ to be the smallest    real number $\delta \in (\epsilon, \frac{1}{2})$ for which  $\text{min-packing}_ {\delta}(d(L))\ne \Phi$. Then, we determine   $c^{*}$ to be an integer in $[\lceil\frac{1}{\epsilon}\rceil,\lceil \frac{2} {\epsilon}\rceil]$ that  minimizes the packing ratio (ie.
$\frac{\text{min-packing}_{\delta_{c^{*}}}(\hat{d}_{c^{*}}(L))}{c^{*}} = 
\min_{c \ \in (1, \lceil \frac{2}{\epsilon} \rceil)}^{} 
\frac{\text{min-packing}_{\delta_{c}}(\hat{d}_{c}(L))}{c}$).
\newline \newline 
{\bf Heuristic C($L$, $\epsilon$)}
\begin{tabbing}
	Input(s): \=  (1) \= $L$ \  = \=  $(a_1,a_2,...,a_n)$ \  be the sequence of $n$ items with  their     respective sizes  \\ 
	\>      \>         \> $(s_1,s_2, ... ,s_n)$ in the interval $[0, 1]$; \\
	\> (2) \> $\epsilon \in (0, \frac{1}{2})$ be a user specified parameter; \\
	Output(s): The assignment of the items in $L$ to the bins in $\cal{B}$; \\
	Begin \= \\ 
	\> (1) \= Let \= $\hat{d}(L) = (s_1*n_1, s_2*n_2, ..., s_k*n_k)$   be the distribution vector corresponding to $L$; \\
	\> (2) \= For \= ($c=1$; $c \le \lceil \frac{2}{\epsilon}\rceil$; $c= c+1$) \\
	\> (2a)    \>     \> Let $\hat{d}_{c}(L) = (s_1*n^{c}_1, s_2*n^{c}_2, ..., s_k * n^{c}_k)$  be the $c$-length segment of $\hat{d}(L)$; \\
	\> (2b)    \>     \> For \= ($\delta=\epsilon$; $\delta \le \frac{1}{2}$; 
	$\delta=\delta+\epsilon$) \\
	\>         \>     \>     \> If $(\text{min-packing}_{\delta}(\hat{d}(L)) \ne \Phi$) break;\\
	\>    (2c)     \>     \>
		Let $\delta_{c} = \delta$ and 	$\text{packing-ratio}(c) = 
		\frac{\text{min-packing}_{\delta_{c}}(\hat{d}_{c}(L))}{c}$; 	\\	
	\> (3)     \>  
		Let $c^{*}$   \=  be an integer in $(\lceil \frac{1}{ \epsilon }\rceil, \lceil \frac{2}{ \epsilon }\rceil)$ : 
		$\text{packing-ratio}(c^{*}) = \min_{c\in (\lceil \frac{1}{\epsilon}\rceil,\lceil \frac{2}{\epsilon}\rceil)}^{} \text{packing-ratio}(c)$; 	\\
	\>  (4) Let $T= \hat{d}_{c^{*}}(L)$  and $l = \frac{\hat{d}(L)}{c^{*}}$; \\
	\>  (5) \> Let $\text{Packing}(\hat{d}(L)) = \bigcup \limits_{i=1}^{l} \text{min-packing}_{\delta_{c^{*}}}(T) \cup \text{min-packing}_{\delta_{c^{*}}}(\hat{d}(L)-l*T)$; \\
	\> (6) \> return $\text{Packing}(\hat{d}(L))$ \\
	End 
\end{tabbing}
{\bf Sub-routine for computing $\text{min-packing}_{\delta}(d_{ c}(L))$}: We first present    a    recurrence that determines $\text{ min-packing}_{\delta}(\hat{d}_{c}(L ))$, for $\delta\in [0,\frac{1}{2}]$, and can be easily converted into a          dynamic program. We then reduce the  computation time of this dynamic  program by presenting a heuristic that  employs the same   recurrence but     restricts   the choice of vectors to a         small sub-set from    $e_{\delta}(L)$. Now, we present some essential  definitions before   presenting   our recurrence and heuristic.
\begin{definitions}
Let $T= (t_1, t_2, ..., t_k)$ denote   $\hat{d}_{c}(L)$, the $c$-length initial segment of $\hat{d}(L)$. Let $\delta\in [\epsilon,  \frac{1}{2} ]$ be a real number and $N = |e_{\delta}(L)|$ denote      the number of $1-\delta$ configurations consistent with $L$. Let $C_1,C_2,..., C_{N}$ denote the complete enumeration of the $1-\delta$        vectors (bins) consistent with   $L$, where $c_{ij}$  denotes the   $i$th component of $C_j$. 
\end{definitions}
From definition, we can observe that $\text{min-packing}_{\delta}(T)$, a minimum sized $\delta$-packing of $T$, is   a  smallest sized collection of vectors from $e_{\delta}(L)$ such that for $i\in [1..k]$, the sum of the $i$th      components of  these vectors   is greater  than or equal to the $i$th component of $T$. So, we can define    $\text{min-packing}_{ \delta}(T)$ recursively as follows:
\begin{equation} 
\label{eqn1}
\text{min-packing}_{\delta}(T) =               \min_{i \in [1..N]} \{ 1 + \text{min-packing}_{\delta}(T-C_i) \}
\end{equation} 
This   recurrence  helps construct $\text{min-packing}_{ \delta}(T)$   by choosing at     most    $\lceil\frac{c}{1-\delta }\rceil$  vectors from $e_{\delta}(L)$ and  can be converted into an $O(N^{\frac{c}{1-\delta}} )$ dynamic program, where $N = |e_{\delta}(L)|\approx(k+\lceil \frac{1} {\delta}\rceil)^{k}$. The computation  time of this dynamic program  is very high, so we design a heuristic that employs the same    recurrence but          restricts the choice of vectors to a small sub-set of size $O(\lceil \frac{c}{1-\delta}\rceil)$ from $e_{\delta}(L)$. \newline \newline
{\bf Key Idea}: For a   given $\delta \in (\epsilon, \frac{1}{2})$ $\text{min-packing}_{\delta}(T)$ is constructed as as follows:
(i) Construct     $e_{\delta}(L)$     efficiently  and store it compactly; (ii) Construct a set $C$ consisting of $O(\lceil     \frac{c}{1- \delta} \rceil)$ vectors randomly chosen from $e_{\delta}(L)$, and (iii) Construct $\text{min-packing}_{\delta}(T)$  using   recurrence Equation $(\ref{eqn1})$ with the choice of vectors restricted to  $C$. \newline \newline 
{\bf Heuristic Packing($T$, $\delta$)}
\begin{tabbing}
	Input(s): \=  (1) \= $T$ a $c$-length segment of $d(L)$, the distribution vector of $L =  (a_1,a_2,...,a_n)$;  \\
	\> (2) \> $\delta \in (0, \frac{1}{2})$ be a user specified parameter. \\
	Output(s): A $\delta$-cover for $T$. \\
	Begin \= \\
	\> (1) \=  Cons\=tructing $e_{\delta}(L)$: \\
	\> (1a)\>      \> Solve the following {\em Knapsack problem(KP(S))}: Given a collection $S$ consisting  \\
	\>     \>      \> of $\lfloor \frac{1} {s_i}\rfloor$ copies of items \ of \ size $s_i$, $i \in [1..k]$,  we \ need to determine the   subsets \\
	\>     \>      \>  of $S$ whose sum is in the interval $[1-\delta, 1]$.\\ 	\>  \>    \> The  standard dynamic programming  solution for $KP(S)$ \ will \ result in a two \\
	\>  \>    \> dimensional table \   consisting \ of \  \ \ $nW$\  entries \ \ \  where \  the\  number\  of \  items \\
	\>   \>    \> $n =    \sum_{i=1}^{k} \lfloor \frac{1}{s_i}   \rfloor$ and  the number of   distinct  weight classifications is   $W = \lceil \frac{1}{\delta} \rceil$. \\ 
	\> (1b) \> \>  Comp\=actly   store $e_{\delta}(L)$  using    a directed graph $KPG(S)$ constructed \ from the\\
	\>      \>  \> dynamic programming table of $KP(S)$: \\
	\>      \>    \>   \> In \ $KPG(S)$ \ there \ are \ $nW$\  nodes with \ each \ node \ associated\  with\  a \\
	\>      \>    \>  \> dynamic \ programming \ entry \  in $KP(S)$.   There\  \ is \ an \ edge \ $(u, v)$ \ in \   \\
	\>      \>    \>   \> $KPG(S)$ if \   the sub-problems in $KP(S)$ corresponding\  to \ nodes $u$ and \\
	\>      \>    \>   \> $v$ are directly related   (i.e.    solution \  to    \   \ sub-problem \ corresponding \ to \\
	\>      \>    \>   \> $v$ can be obtained from the  sub-problem\  corresponding \ to   $u$   by   adding \\
	\>      \>    \>   \> a \ single\  item\  in $S$), \ and \ the 
	weight associated\  with \ the edge  $(u, v)$ is     \\
		\>      \>    \>   \> the weight of the item \ that relates these two sub-problems. \\ 
		\>	    \>    \>  Note: \= There \ is \ a \ $1-1$ \ correspondence \ between \ the \ $nW$ sub-problems in \\
	\>      \>     \>   \> $KP(S)$\  and \ the\  $nW$\  table \ entries of a  dynamic  programming  solution \\
	\> \> \> \> for $KP(S)$.\\ 
		\> (2) \=  Cons\=tructing  \ $C$ \ from \ $e_{\delta}(L)$: \\ 
	\> (2a) \> \> Construct \ \ $KPG'(S)$   by removing    \ all
	 \ useless  \ nodes \ from $KPG(S)$, where \\
	\> \> \>       a   \ node $(i, w)$ is {\em useless} \ if  \ its  weight is \ in  the interval $[0,1-\delta)$ \ and has no \\
	\> \>  \>  directed edge to a node  with greater weight.
	\\
	\> (2b) \> \> Cons\=truct \ the \ set $C=(S_1,S_2,...,S_l)$ of size $\lceil\frac{c}{1-\delta}\rceil$ \ by  choosing $S_k$, $k \in
	[1..l]$     \\
	\>     \> \> as follows: \\
	\>     \>  \>  \> Set the current node to $s$;   while   the current \ node $i$ \ has \ an \ out degree \\
	\>     \>   \> \> greater than $0$, \ choose  uniformly at  random  a    \ directed \ edge $(i, j)$\  from \\
	\>     \>   \> \> among the edges leaving  node $i$ and  include the weight corresponding to \\
	\>     \>  \> \> that edge $(i, j)$ in set $S_k$. Now set current node to node $j$ and repeat the\\
	\>     \>   \> \> above step.  \\
	\> \> \> Note: \=  There is a $one-one$ correspondence between the 
	vector constructed \ by \\
	\> \> \> \> following \ a \ directed \ path\  from \ $(0, 0)$ to $(i, w)$ in $KPG(S)$, \ where  $i \in $ \\
	\> \> \> \> $[0,  n]$ and  $w \in [1-\delta, 1]$, \ and \ a vector in $e_{\delta}(L)$.   So, we \ construct \ $C$ by \\ 
	\> \> \> \>  sampling uniformly \ at \ random from paths in $KPG(S)$ that correspond   \\
	\> \> \> \> to vectors in  $e_{\delta}(L)$. In \ $KPG(S)$ \ since there are directed paths  \ that  do   \\
	\> \>  \> \>  not correspond to \ a \ vector \  in \ $e_{\delta}(L)$, we \ 
	\ modify \ $KPG(S)$ to \ obtain   \\
	\>  \> \>     \>   $KPG'(S)$ where  \  there \   is \ an \ $1-1$ correspondence between \ a  directed \\  
	\> \> \> \> \  path\  from node $s=(0, 0)$  \ to a node with out degree  $0$,   and \ a vector in  \\
	\> \> \> \> $e_{\delta}(L)$. \\
	\> (3) Construct \= $\text{min-cover}_{\delta}(T)$ by modifying the  Equation $(\ref{eqn1})$ as follows: \\
	\> \>     $\text{min-cover}_{\delta}(T) =     \min_{i \in [1..\lceil\frac{c}{1- \delta} \rceil]} \{ 1 + \text{min-cover}_{\delta}(T-S_i) \}$ \\
End 
\end{tabbing}
\section{Empirical Analysis of Heuristic $C$}
In this section we present        our empirical study    of Heuristic $C$ from two perspectives: 
(i) solution quality - the nature of approximation           guarantees it can provide; and (ii) computational efficiency - scalability of the heuristic. We desire an algorithm that can provide  near optimal solutions and is computationally efficient. However, there    is a natural tradeoff  between solution quality and computational efficiency that is made worse by the      hardness of the bin packing problem. Our Heuristic $C(L,\epsilon)$, where $\epsilon\in (0,1)$ is the desired error bound, attempts to obtain an $1+\epsilon$ optimal solution by essentially     breaking the original problem into many identical sub-problems of size $c\in    [1,\frac{2}{1-\epsilon}]$ and  then  solving that sub-problem using at    most $N  \in [1,\frac{c}{1- \epsilon}]$ distinct  bin configurations. For most instances Heuristic $C$       splits the original problem instance into identical      sub-problems of size  $c < 10$ which is then solved by considering less than $10$  distinct  bin configurations (i.e. $N < 10$). For some instances Heuristic $C$ needs to consider $\approx 25$ distinct bin           configurations in order to satisfy the performance guarantee constraint.     For     these instances Heuristic $C$ is  scalable and provides   good guarantee on its solution quality. For a  very small fraction of instances  Heuristic $C$ does not provide the    approximation guarantee when analyzed using traditional means. However, for most of these instances we were able to  obtain the desired performance guarantee    by     obtaining a better lower bound on the necessary wastage in an optimal solution. We  now present our empirical study of Heuristic $C$ by first describing our experimental set-up and experiments, and then presenting  the experimental results      and our observations. \newline \newline
{\bf Experimental setup}: We created two sets of sequences: (i) {\em Sequence-Set-H}: A set of $300$ instances obtained by randomly partitioning a unit interval into triplets,   quartets or quadruplets. These are  combinatorially hard instances for which we know an   optimal solution with wastage almost zero and hence for these hard instances our experimental analysis is tight
; (ii) {\em Sequence-Set-R}: A set of $1000$ instances where the item sizes are drawn randomly from a distribution   parameterized by the number of item types. These are very few  instances 
for which we do not necessarily know the optimal and also we do not have a good lower bound on 
the wastage in an optimal solution. Hence traditional analysis for these instances may not  be tight. We now describe how we generate $Sequence-Set-H$ and $Sequence-Set-R$.\newline \newline
{\bf Sequence-Set-H}: For each $l\in [3..5]$, we use $Generate-h(n, l)$ to generate a   random sequence $L(n,l)$ of length $nl$ obtained by randomly partitioning $n$ unit intervals into $l$ pieces each. We  generate $100$ such sequences for each value of   $l$ as follows:
 \begin{quote} 
{\bf $Generate-h(n, l)$}: Create $l$-items by  randomly     partitioning the    unit  interval $(0,1)$ into $l$ pieces by using $l-1$ cut points drawn    from standard  uniform distribution and            rounded to the nearest    multiple of $0.05$  as   cut points, and then use the  lengths of   these   $l$ pieces to be the sizes of the $l$ pieces obtained by partitioning the unit interval. Repeat this step $n$ times.
\end{quote} 
{\bf Sequence-Set-R}:  For    $n = 1000$ and   each $k \in [6..15]$, we use $Generate-r(n, k)$ to   generate   a random     sequence $L(n,k)$ of length   $n$ consisting of       at most $k$ distinct item sizes. We generate $100$ such sequences for      each value of   $k$ as follows:
\begin{quote} 
{\bf $Generate-r(n, k)$}: First, determine         the $k$ item sizes $\{s_1,s_2,...,s_k\}$ by generating a sample of size $k$ where each item is drawn from a standard  uniform distribution  and rounded to the nearest multiple of $0.05$; Second, partition  the unit interval $(0,1)$ by using           $k-1$ cut points drawn from standard uniform distribution, and then use    the  lengths of the $k$ pieces obtained by scanning the unit interval from left to right to specify $(p_1, p_2, ..., p_k )$, the probability distribution of item sizes in  $L(n, k)$; and finally generate $L(k, n)$ by simulating a multinomial distribution using   $(p_1, p_2, ..., p_k )$. 
\end{quote} 	
{\bf Note}: The sequences generated using $Generate-h$  are similar to the   instances used by Falkenauer (i.e. Falkenauer Triplets). Here we generate triplets, quadruplets and quintuplets. 
We refer the reader to BPPLib \cite{DIM18} for an excellent and comprehensive    collection of codes, benchmarks, and links for the    one-dimensional Bin Packing and Cutting Stock problem.
\newline \newline 
{\bf Experiments}:
We ran       Heuristic $C$ on instances in $Sequence-Set-H$ and Heuristic $C$, $BFD$ (Best Fit Decreasing) and   $FFD$ (First Fit Decreasing) on instances in $Sequence-Set-R$. For instances 
in $Sequence-Set-H$, we    wanted an $(1 +\epsilon)$-optimal solution, where $\epsilon = 0.1$. 
For instances in $Sequence-Set-H$, we    wanted   a solution whose quality  is better than the    solutions obtained through either $BFD$ (Best Fit Decreasing) or $FFD$ (First Fit Decreasing).  For     each sequence, we observed the    following: (i) $c$ - the size of the sub-problem  it partitions         the     input instance into; (ii) $N$ - the number of bin configurations it considers while solving the   $c$-sized    sub-problem in (i); and (iii) lower bound    on the necessary wastage of an optimal solution for the instances  in $Sequence-Set-R$ where   we are 
not able to guarantee $1+\epsilon$ optimality.   \newline \newline 
{\bf Experimental Results}:    For instances in $Sequence-Set-H$, $k$ ranged from $3$ to $20$. However, for $\approx 80\%$ of the instances $k$ is in $[8..16]$. For instances where $k < 5$, Heuristic $C$ is    able to get the desired quality for $c < 10$ and $N < 10$. Also, for   $k >10$,   Heuristic $C$ is able to get the desired quality for $c < 10$ and $N < 15$.  However, 
for $k \in [5..9]$, there  are some instances where we are unable to get the desired solution quality for $N < 25$ (irrespective of the value of $c$). For these instances, the performance 
is very sensitive to the heuristic's choice of configurations. So for the randomly chosen bin 
configurations to contain some specific collection of bin configurations our heuristic ends up picking a larger sample. 
For $Sequences-R$, for instances  where $k < 7$ and $k > 10$, we are able to get the solution 
of desired  quality for $c < 10$ and $N < 10$.However, for    some instances where     $k \in [7..10]$ and the item size distribution is skewed to the right (i.e. many items of size  $\ge 
0.4$) Heuristic $C$ is able to perform as good as the best of $BFD$ and $FFD$ but traditional 
analysis is unable to provide guarantee about its near optimality mostly due to the inability 
to get a good lower bound on the necessary  wastage in any optimal solution. In most of these
instances we are able to improve the lower bound and hence the performance guarantee.
 \newline \newline 
{\bf Conclusions}: We are able to design a simple heuristic that our    preliminary empirical 
study indicates is highly scalable and is amenable to tighter analysis due to the use of bins 
with wastage as close to $\epsilon$ as possible. For     most instances   it is able to scale because      it is able to split the given sequence of $n \approx 1000$ items  into identical sub-problems of length $c$ for $c \in [10, 20]$ and each  of these $c$-length sub-problems is solved using fewer than    $10$ distinct bin configurations for most instances. However, when 
$k$- the number of item sizes $k$ is in $[7..10]$  and   the average  item size is $>0.4$ our heuristic is not  able to guarantee       near optimality for some very few instances  partly     because of sensitivity of the instance to the choice of bin configurations and mostly due  to
the inability to get a good lower bound on the necessary  wastage in any optimal solution for these instances.

\end{document}